# Extensible Generic Data Management Software
Workshop on Sustainable Software for Science: Practice and Experiences
Reagan Moore
University of North Carolina at Chapel Hill

Extensibility mechanisms constitute a form of knowledge capture that is essential for software re-use.  The Data Intensive Cyber Environments (DICE) group has collaborated with more than 25 science and engineering domains on the design of open source distributed data management systems.  A major requirement has been the provision of extensibility mechanisms that enable provision of domain-specific research capabilities. A partial list of domains and projects that build on extensible generic data management infrastructure includes:

| | |
|---|---|
| Astrophysics | Auger supernova search |
| Atmospheric science | NASA Langley Atmospheric Sciences Center |
| Biology | Phylogenetics at CC IN2P3 |
| Climate | NOAA National Climatic Data Center |
| Cognitive Science | Temporal Dynamics of Learning Center |
| Computer Science | GENI experimental network |
| Cosmic Ray | AMS experiment on the International Space Station |
| Dark Matter Physics | Edelweiss II |
| Earth Science | NASA Center for Climate Simulations |
| Ecology | CEED Caveat Emptor Ecological Data |
| Engineering | CIBER-U |
| High Energy Physics | BaBar / Stanford Linear Accelerator |
| Hydrology | Institute for the Environment, UNC-CH; Hydroshare |
| Genomics | Broad Institute, Wellcome Trust Sanger Institute, NGS |
| Medicine | Sick Kids Hospital |
| Neuroscience | International Neuroinformatics Coordinating Facility |
| Neutrino Physics | T2K and dChooz neutrino experiments |
| Oceanography | Ocean Observatories Initiative |
| Optical Astronomy | National Optical Astronomy Observatory |
| Particle Physics | Indra multi-detector collaboration at IN2P3 |
| Plant genetics | the iPlant Collaborative |
| Quantum Chromodynamics | IN2P3 |
| Radio Astronomy | Cyber Square Kilometer Array, TREND, BAOradio |
| Seismology | Southern California Earthquake Center |
| Social Science | Odum, TerraPop |

Each domain has unique semantics, data formats, types of data, analysis procedures, management policies, descriptive metadata, and hardware systems (including unique network access protocols).  Each domain has existing infrastructure that manages legacy data, provides analysis services, and serves as an authoritative resource for domain knowledge.  Each domain may organize data in a collection, or share in a data grid, or publish in a digital library, or preserve in an archive.

We claim that extensible software enables infrastructure re-use across domains.

The approach taken in the DICE group has been to build middleware that captures domain knowledge.  This may be knowledge that is need to translate from access protocols required by domain resources to the client protocols desired by researchers.  Or it may be knowledge needed to provide a unified view across heterogeneous production systems.  The unified view constitutes a collaboration environment through which researchers can access existing resources, share data, information, and knowledge, and manage their research data.  The collaboration environment provides infrastructure independence, enabling re-use of existing hardware and software systems.

Essential components of domain knowledge are captured in mechanisms that enable infrastructure independence.  The domain knowledge is captured in interoperability mechanisms that enable use of multiple types of technology within the collaboration environment.  The interoperability mechanisms can be categorized through the types of operations that each domain performs upon the following seven name spaces:
- Users        (group formation, authorization, authentication, audit)
- Resources   (storage interaction, remote application execution, queuing)
- Files          (replication, versioning, distribution, streaming, transport)
- Collections  (access controls, archiving, soft links, registration)
- Metadata    (schema, ontologies, vocabularies)
- Policies      (enforcement points, automation, versioning)
- Procedures  (workflow provenance, re-execution, versioning, sharing)

The seven name spaces have been implemented in the iRODS integrated Rule Oriented Data System, along with the interoperability mechanisms that capture the knowledge needed to apply the desired operations across existing hardware and software systems.  Note that the approach has to enforce management policies across administrative domains, provide a single sign-on environment for users, enable re-use of existing data collections, enable processing both at the place where data are stored and at compute engines, and maintain a consistent and persistent set of provenance, descriptive, and administrative metadata.

Extensibility mechanisms represent forms of knowledge capture.  The knowledge process that should be applied to access a remote system or execute a procedure is captured within an interoperability mechanism.  Sustainable software provides the interoperability mechanisms needed to incorporate new technology.  The approach taken for building sustainable software is best illustrated through examples of requirements from user communities, and through a description of the generic knowledge capture mechanisms that were implemented to meet each requirement.

A dominant requirement has been the ability to capture management knowledge in computer actionable rules.  A driving requirement from the UK e-Science data grid was a request for the ability to create a collection in which files were permanently managed and could never be deleted by anyone.  But at the same time, the ability to manage a collection in which administrators could replace corrupted files was desired, and the ability for users to update their own files in their own collections.  This implied the need to manage at least three different consistency constraints on data deletion within the same data management system (no deletion allowed, deletion by administrator, deletion by file owner).

The DICE group developed the iRODS policy-based system to extract knowledge about management policies from the software, and apply the knowledge through computer actionable rules stored in a rule base. Effectively, every software encoded consistency constraint was replaced by a policy-enforcement-point. Actions by clients were trapped at the policy-enforcement-points. By searching the rule base, an appropriate rule could then be identified which controlled the execution of a workflow that enforced the required management policy. This meant that the knowledge needed to manage the system could be captured in computer actionable rules. The system was no longer restricted to managing files and static representations of information. Instead, a data management system could use rules that controlled the behavior of the system and administrators could dynamically change the rules in a rule base. It became possible to use generic infrastructure to implement archives, digital libraries, data grids for sharing data, project collections, and processing pipelines simply by changing the rules and procedures enforced by the system.

Within iRODS, policies can be enforced for preservation (authenticity, integrity, chain of custody, original arrangement, retention, disposition); or for data publication in a digital library (descriptive metadata annotation, arrangement, creation of presentation versions such as image thumbnails); or for sharing in a data grid (access controls, distribution, caching); or for reproducible data driven research in a processing pipeline (workflow procedures, workflow provenance, workflow re-execution); or for validating assessment criteria (repository trustworthiness, compliance with regulations).

A second form of knowledge capture is the management of provenance and descriptive metadata. Each science and engineering domain uses different descriptive terms. The iRODS data management system uses schema indirection to enable each community to apply their desired metadata. In essence, descriptive and provenance information are turned into triplets: metadata attribute name; metadata attribute value; and metadata attribute comment. This approach makes it possible for each community to independently specify the information context associated with their collections.

A third form of knowledge capture is the automated management of workflow provenance information. Within iRODS, a workflow collection can be associated with a file that contains a workflow written in a workflow language. The output from each execution of the workflow can be captured, along with the input files. This enables reproducible data-driven research through the sharing of the workflow, the input files, and the output files. A researcher can re-execute an analysis done by another scientist, modify the input files, re-run the analysis and compare results.

A fourth form of knowledge capture is the automated management of data streams. Within iRODS, an archive collection can be associated with a data stream. Data that are deposited into the archive collection are automatically indexed based on a stream time parameter. Data within a specified time interval can then be retrieved in a single data stream. The data grid automatically does the required sub-setting of files for the start and end of the stream, and composes the intermediate files into the requested data stream.

A fifth form of knowledge encapsulation is a basic function (micro-service) that manages the network protocol needed to interact with an external data repository. The micro-service manages the communication, and caches the retrieved data within the collaboration environment. This enables a researcher to link external data sets into a collaboration environment, apply analyses, and manage results while maintaining control over the input files.

Each of these types of knowledge capture is an example of re-use of data grid infrastructure to support a new science and engineering domain. Based on experience with 25 domains, three types of interoperability mechanisms are needed to capture domain knowledge needed for software re-use:

1. Policies that control the execution of procedures, management of data, and verification of assessment criteria.
2. Micro-services that manage interactions with external network protocols, encapsulate specific operations, and encapsulate workflow operators (conditional tests, loops, arithmetic).
3. Drivers that apply data grid operations at remote storage locations (Posix I/O commands, staging, archiving).

Using these three interoperability mechanisms, the iRODS software has been successfully re-used for projects ranging from institutional repositories, to regional data grids, to national data grids and national libraries, to international collaborations.

Re-use of software is also facilitated through the ability to dynamically change data management system components. In the E-iRODS software, each of the interoperability mechanisms is dynamically pluggable. It is possible to add a new policy, add a new micro-service, and add a new storage driver while the system is running. This makes it possible to evolve production environments. New resources can be added without having to stop the production system.

Today, viable data management systems automate enforcement of management policies, automate administrative tasks such as data migration, automate validation of assessment criteria, capture knowledge (processes) associated with creating derived data products, capture knowledge (communication protocols) needed to interact with remote systems, and automate processing of data within workflow pipelines. The automation of these tasks corresponds to the creation of knowledge procedures that can be applied by a policy-based data management system.

iRODS:  https://www.irods.org

E-iRODS: http://www.e-irods.org